\documentstyle[11pt,aaspp4,psfig]{article}
\def\sso{$^{-1}$}
\def\kms{km\ s$^{-1}$}

\def\Msun{$M_{\sun}$}

\def\Ha{H$\alpha$\ }
\def\bm {beam$^{-1}$}
\def\yr {yr$^{-1}$}
\def\ecs {ergs\ cm$^{-2}$\ s$^{-1}$}
\received{}
\accepted{}
\journalid{}{}
\articleid{}{}
\slugcomment{}

\lefthead{Sheth et al.}
\righthead{Molecular Gas, Dust and Star Formation in NGC~5383}

\begin{document}
\renewcommand{\topfraction}{1}
\renewcommand{\textfraction}{0.1}
\renewcommand{\floatpagefraction}{0.1}

\title{Molecular Gas, Dust and Star Formation in the Barred Spiral
  NGC~5383\footnote{Based on observations with the NASA/ESA Hubble Space Telescope, obtained at the Space Telescope Science Institute, which is operated by the Association of Universities for Research in Astronomy, Inc. under NASA contract No. NAS5-26555.}}
 
\author{Kartik Sheth\altaffilmark{1}}
\affil{Astronomy Department, University of Maryland, College Park, MD 20742-2421 \\}

\author{Michael W. Regan\altaffilmark{2,3,4}}
\affil{Carnegie Institution of Washington, Department of Terrestrial
  Magnetism, 5241 Broad Branch Road, Washington D.C. 20015}

\and

\author{Stuart N. Vogel\altaffilmark{5}, \& Peter J. Teuben\altaffilmark{6}}
\affil{Astronomy Department, University of Maryland, College Park, MD 20742-2421 \\}

\altaffiltext{1}{E-mail: kartik@astro.umd.edu}
\altaffiltext{2}{E-mail: mregan@dtm.ciw.edu}
\altaffiltext{3}{Hubble Fellow}
\altaffiltext{4}{Visiting Astronomer, Kitt Peak National Observatory,
  National Optical Astronomy Observatories, which is operated by the
  Association of Universities for Research in Astronomy, Inc. (AURA)
  under cooperative agreement with the National Science Foundation.} 
\altaffiltext{5}{E-mail: vogel@astro.umd.edu}
\altaffiltext{6}{E-mail: teuben@astro.umd.edu}
\clearpage

\begin{abstract}
  We have mapped the barred spiral NGC 5383 using the BIMA
  millimeter-wave array for observations of CO (J=1--0), the Palomar
  1.5m for H$\alpha$\ and optical broadband, and the Kitt Peak 1.3m
  for near-IR broadband.  We compare the observed central gas and dust
  morphology to the predictions of recent hydrodynamic simulations
  calculated using the Piner, Stone, and Teuben code.  In the nuclear
  region, our observations reveal three peaks lying along a S-shaped
  gas and dust distribution: two of these are at the inner end of
  offset bar dust lanes at the presumed location of the inner Lindblad
  resonance, and the other lies closer to the nucleus.  In contrast,
  the model predicts a circumnuclear ring, not the observed S-shaped
  distribution; moreover, the predicted surface density contrast
  between the central gas accumulation and the bar dust lanes is an
  order of magnitude larger than observed.

  These discrepancies remain for all our simulations which produce
  offset bar dust lanes, and indicate that the model is missing an
  essential process or component.  A small nuclear bar might account
  for the discrepancy, but we rule this out using a HST NICMOS image:
  this reveals a nuclear trailing spiral, not a bar; we show that
  coarser resolution (i.e. ground-based images) can produce artifacts
  that resemble bars or rings.  We conclude that the discrepancies in
  morphology and contrast are due to the omission of star formation
  from the model; this is supported by the observed high rate of
  central star formation (7 \Msun\ \yr), a rate that can consume most
  of the accumulating gas.

  As is common in similar bars, the star formation rate in the bar
  between the bar ends and the central region is low (~0.5 \Msun\ 
  \yr), despite the high gas column density in the bar dust lanes;
  this is generally attributed to shear and shocks.  We note a
  tendency for the HII regions to be associated with the spurs feeding
  the main bar dust lanes, but these are located on the leading side
  of the bar.  We propose that stars form in the spurs, which provide
  a high column density but low shear environment.  HII regions can
  therefore be found even on the leading side of the bar because the
  ionizing stars pass ballistically through the dust lane.

\end{abstract}


\clearpage
\section{INTRODUCTION}

Barred spirals constitute a large fraction of all disk galaxies
($\sim$ 75\% according to a near-IR survey by
\markcite{mulchaey97a}Mulchaey, Regan \& Kundu 1997).  
The non-axisymmetry of a bar potential induces gas inflow which can
lead to a number of dramatic evolutionary changes in a galaxy.  These
include large central concentrations of molecular gas
(\markcite{kenney92}Kenney et al. 1992; \markcite{kenney96}1996 and
references therein; \markcite{sakamoto99}Sakamoto 1999), central
starburst activity (\markcite{heller94}Heller \& Shlosman 
1994; \markcite{phillips93}Phillips 1993;\markcite{garcia96} 
Garcia-Barretto et al. 1996; \markcite{ho97}Ho et al. 1997),
reduction of the overall chemical abundance gradient
(\markcite{martin94}Martin \& Roy 1994; \markcite{martinet97}Martinet
\& Friedli 1997), formation of a new bulge (\markcite{norman96}Norman,
Sellwood \& Hasan 1996), transfer of angular momentum to the halo via
dynamical friction (\markcite{tremaine84}Tremaine \& Weinberg 1984;
\markcite{weinberg85}Weinberg 1985) and the destruction of the bar
itself (\markcite{norman96}Norman et al. 1996;
\markcite{sellwood99}Sellwood \& Moore 1999).  

\begin{deluxetable}{lrr}
\small
\tablecaption{Properties of NGC~5383 \label{global}}
\tablehead{\colhead{Parameter}&\colhead{Value}&\colhead{Reference}}
\startdata
R.A. (J2000)\tablenotemark{a} & 13$^h$57$^m$04$^s$.81 & (1) \nl
Dec (J2000)\tablenotemark{a}  & 41$^d$50$^m$47$^s$.68 & (1) \nl
D$_{25}$ & 2$\farcm$75 & (2) \nl
Major axis P.A. & 85$\pm$3$^o$ & (2) \nl
Inclination & 50$^o$ & (2) \nl
V$_{sys}$ & 2250 \kms~ & (3) \nl
V$_{3K}$ & 2428 \kms~ & ... \nl
Adopted Distance & 32.4 Mpc & ... \nl
Linear scale & 157 pc arcsec$^{-1}$ & ... \nl
Bar Length & 110$\pm$3$\arcsec$ & (1) \nl
Bar P.A. & 130$\pm$4$^o$ & (1) \nl
\tablenotetext{a}{K$^\prime$  band peak}
\tablerefs{(1) This paper (2) Duval \& Athanassoula 1983 (3) Becker,
  White, \& Helfand 1995.} 
\enddata
\end{deluxetable}

Though a majority of these effects depend on the gas inflow, there is
no consensus on the exact mechanism for the inflow.  There are at
least two classes of models: cloud-based/sticky particle simulations
which treat the gas as a collection of discrete particles subject to
``sticky'' collisions \markcite{combes85}(Combes \& Gerin 1985) and
hydrodynamic simulations which treat gas as an ideal fluid
(\markcite{a92b}Athanassoula 1992b and references therein;
\markcite{piner95}Piner, Stone \& Teuben 1995, hereafter PST95).  Both
models form dust lanes along the leading edge of the bar, have mass
inflow, and form circumnuclear rings (\markcite{combes85} Combes \&
Gerin 1985; \markcite{piner95}PST95).

The hydrodynamic models have been remarkably successful in explaining
the shapes of the main bar dust lanes, the shock signature at the
location of the dust lane (\markcite{a92b}Athanassoula 1992), and the
overall gas kinematics in the bar (\markcite{regan97}RVT97).
Therefore we use the framework provided by this class of models to ask
two important questions: Can the nuclear gas and dust morphology in
bars be explained by these models?  And can the gas flow predicted by
these models provide a basis for explaining the star formation
activity in the bar?

To answer these questions, we use a multi-wavelength dataset for the
barred spiral NGC 5383; the dataset includes maps of millimeter wave
CO (J=1--0) emission to trace the molecular gas, \Ha emission to trace
the ionized component, and broad-band optical and near-IR emission to
trace the stellar distribution and dust extinction.  NGC~5383 is a
prototypical early Hubble type (SBb) barred spiral galaxy.  The global
characteristics of NGC~5383 are summarized in Table \ref{global}.
NGC~5383 has been studied with optical images
(\markcite{burbidge62}Burbidge, Burbidge \& Prendergast 1962;
\markcite{duval83}Duval \& Athanassoula 1983;
\markcite{elmelm85}Elmegreen \& Elmegreen 1985), long-slit spectra
(\markcite{peterson78}Peterson et al. 1978; \markcite{duval83}Duval \&
Athanassoula, 1983), HI interferometric observations
(\markcite{sancisi79}Sancisi, Allen \& Sullivan 1979), and single-dish
CO observations (\markcite{ohta86}Ohta, Sasaki \& Saito 1986).  The
standard model for some of the seminal studies of gas flow in barred
spirals has also used NGC~5383 as the prototype
(\markcite{huntley78}Huntley 1978;\markcite{sanders80} Sanders and
Tubbs 1980; \markcite{tubbs82}Tubbs 1982; \markcite{duval83}Duval \&
Athanassoula 1983;\markcite{a92b} A92).

We first present a comparison of nuclear gas and dust morphology in
NGC~5383 to the model predictions in \S \ref{hydromorph}.  The gas and
dust distribution is a signature of the bar induced inflow and its
subsequent evolution.  If the hydrodynamic model can reproduce the
observations, then it may be used as an adequate model for studying
gas flow in the nuclear region of bars.  On the other hand, even a
failure of the model can refine our understanding by pointing out
missing physics in the model's assumptions.  We find that while the
model is consistent with the observations in some aspects, there are
several striking differences between the two.  Possible reasons for
these discrepancies are investigated in \S \ref{understand} where we
consider presence of a nuclear bar, incomplete parameter space
exploration of the model, or lack of star formation processes in the
model.  We conclude that the latter is at least partly responsible for
the discrepancies.

In \S \ref{stardust}, we compare the sites of star formation with the
molecular gas distribution in the bar and investigate where and under
what circumstances stars form in the bar.  This is important because
bar star formation can affect the net mass inflow.  The analysis is
novel because previous studies (\markcite{tubbs82}Tubbs 1982;
\markcite{a92b}Athanassoula 1992b; \markcite{reynaud98}Reynaud \&
Downes 1998) have only sought to explain the lack of bar star
formation.  We present the first explanation of how stars may be
forming in the bar, between the bar ends and the circumnuclear
region.  

\section{OBSERVATIONS AND DATA REDUCTIONS} \label{data} 

\subsection{Optical Observations} 
We observed NGC~5383 in a broad band ($\Delta$ $\lambda$ $\sim$ 1240
\AA) R-filter and a narrow band ($\Delta$ $\lambda$ =  25 \AA) \Ha
filter at the 1.5m telescope at Palomar\footnote{Observations were
  made on the 60 inch telescope at Palomar Mountain, which is jointly
  operated by the California Institute of Technology and the Carnegie
  Institution of Washington.} 
on the night of 2 June, 1994 
with a 2048 $\times$ 2048 CCD camera in direct imaging mode.
In this mode the camera has a 12$\farcm$7 field of view with 0$\farcs$37
pixels. 
Using the R-band filter we obtained one 300 s and one 600 s
exposures which were co-added to obtain a single R-band image.  
The \Ha image was a single 900 s exposure.
The effective resolution was 1$\farcs$5 in the R-band image and 2$\farcs$4
in the \Ha image.    
For both images we divided by a normalized flat-field frame,
subtracted the mean sky brightness, and removed cosmic rays using
standard routines in the NOAO/IRAF\footnote{IRAF is distributed by the
  National Optical Astronomy Observatories, which are operated by the
  Association of Universities for Research in Astronomy, Inc., under
  cooperative agreement with the National Science Foundation} software
package.

\begin{figure}[!htb]
\centerline{\psfig{figure=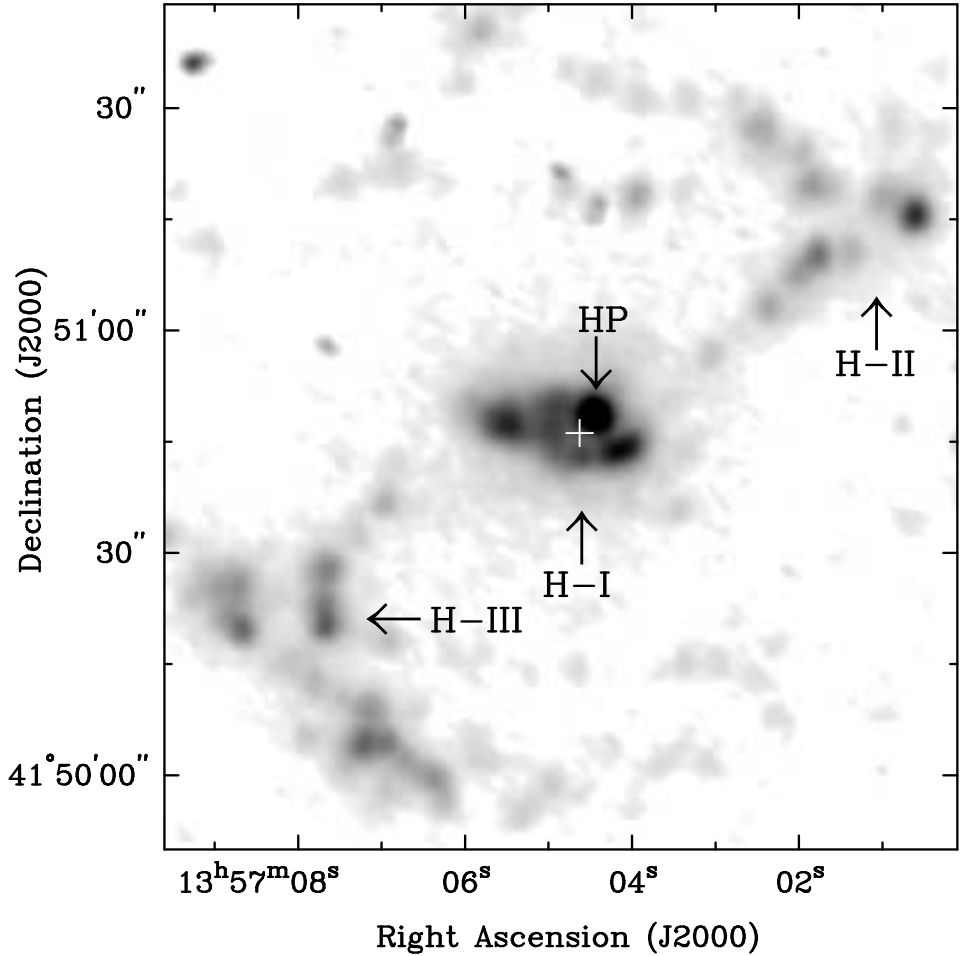,angle=-90,silent=1,width=5in}}
\caption[Continuum-subtracted \Ha image of NGC~5383]
{Continuum-subtracted \Ha image of NGC~5383 showing active
  star-forming regions.  Note the relatively weak \Ha emission along
  much of the length of the bar compared to the strong emission in the
  nucleus and the bar ends.  The peak \Ha
  emission is indicated by the label HP and is displaced 3$\farcs$1
  north northwest from the radio continuum center (indicated by the
  cross).  The labels H-I (nuclear region), H-II (northwestern bar end
  + spiral arm) and H-III (southeastern bar end + spiral arm) indicate
  the three regions where star formation rates were
  calculated.\label{hafig}}
\end{figure}

We also corrected all images for atmospheric extinction.
We used the standard stars BD+26.2606 and BD+17.4703 for absolute flux
calibration using the absolute photometry by \markcite{oke83}Oke \&
Gunn (1983).  
Finally, we registered foreground stars in each image with the Hubble
Guide Star Catalog (GSC) and determined astrometric solutions for each
image.
The residuals in determining absolute positions are smaller than
0$\farcs$03; however, systematic errors in the GSC prevent us from
achieving an accuracy $<$ 1$\arcsec$. 
We used the R-band image to subtract the underlying continuum from the
\Ha image.   
The final continuum-subtracted \Ha line image is shown in Figure \ref{hafig}.

\subsection{Infrared Observations}

\begin{figure}[!htb]
\centerline{\psfig{figure=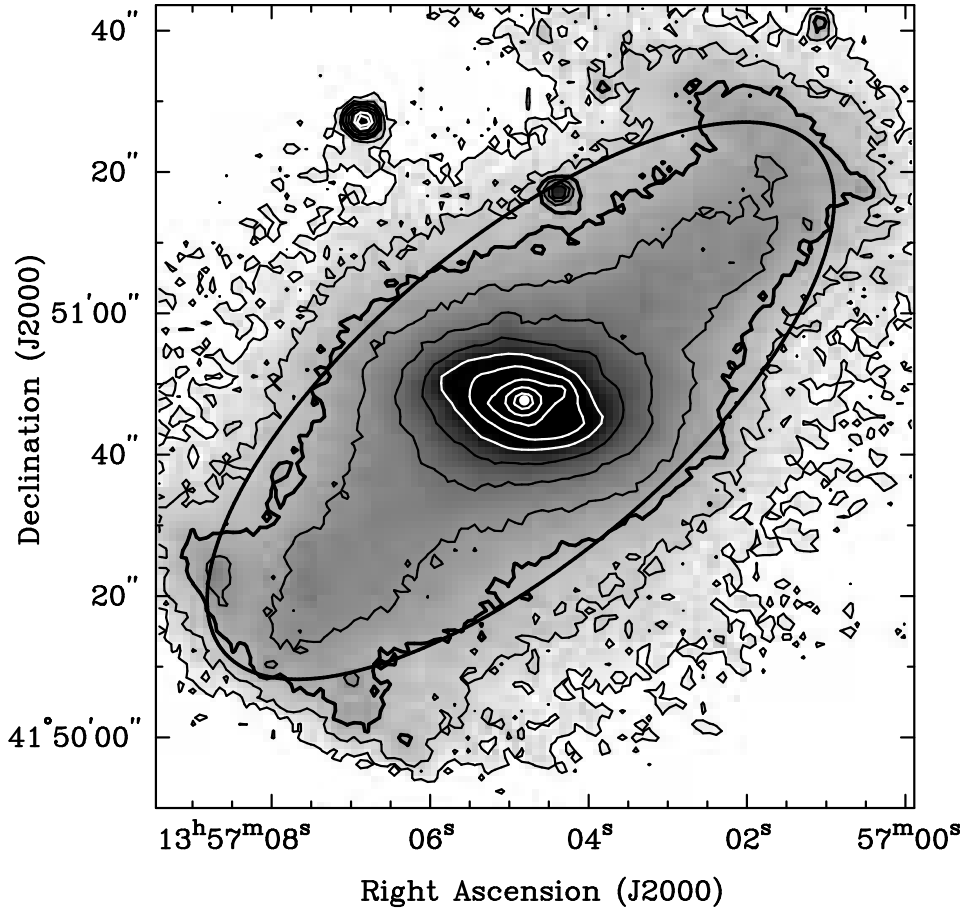,angle=-90,width=5in}}
\caption[K-band (2.25$\mu$m) image of NGC~5383 ]
{K-band (2.25$\mu$m) image of NGC~5383 tracing the old stellar
  population depicts the relatively smooth bar potential.  The
  contours are at 20, 19.5, 19, and in steps of 0.5 magnitudes
  arcsec$^{-2}$ thereafter.  The thick ellipse is the ellipse fit with
  the highest ellipticity and therefore it identifies the bar. The
  thick contour corresponds to the mean intensity along the best-fit
  ellipse. \label{kbandfig}}  
\end{figure}

We observed NGC~5383 in the near-infrared J and K$^\prime$  bands
\markcite{wainscoat92}(Wainscoat \& Cowie 1992) on the nights of 1, 4
March 1994 at the 1.3m telescope on Kitt Peak using the Cryogenic 
Optical Bench (COB). 
The COB used a 256 $\times$ 256 InSb detector with 0$\farcs$95 pixels
($\sim$4$\arcmin$ field of view).
The total on-source integration time was 24 minutes in the J band and
52 minutes in the K$^\prime$  band. 
The effective resolution is 2$\farcs$4 in the K$^\prime$ band image and
2$\arcsec$ in the J band image.
The data were reduced via the procedure described by
\markcite{regan95}Regan et al. (1995). 
Since the K$^\prime$ band image traces light primarily from the old
stars which are good tracers of the gravitational potential, Figure
\ref{kbandfig} is shown to emphasize the bar potential in NGC~5383.  

\subsection{NICMOS Observations}

We observed the nucleus of NGC~5383 with the F160W filter on camera 2
of the Near Infrared Camera and Multi-Object Spectrometer (NICMOS) in
the Hubble Space Telescope on 13 October 1997. 
The F160W filter is an approximate match to ground-based near infrared 
H-band filters and has a central wavelength of 1.6\micron.
Our total integration time on source was 704 seconds.
The detector on Camera 2 is a 256$\times$256 HgCdTe detector with a
plate scale of 0.075 arc seconds pixel\sso\ yielding a field of view of
19\farcs2.
During the time period of the observations the pipeline data reduction
for NICMOS was still under development leading to output of the
pipeline process that was not the best possible calibration of the data.
Therefore, we re-calibrated the images using the latest flat fields, darks,
and non-linearity files. 

\subsection{ BIMA CO Interferometric Observations}

\begin{figure}[!htb]
\centerline{\psfig{figure=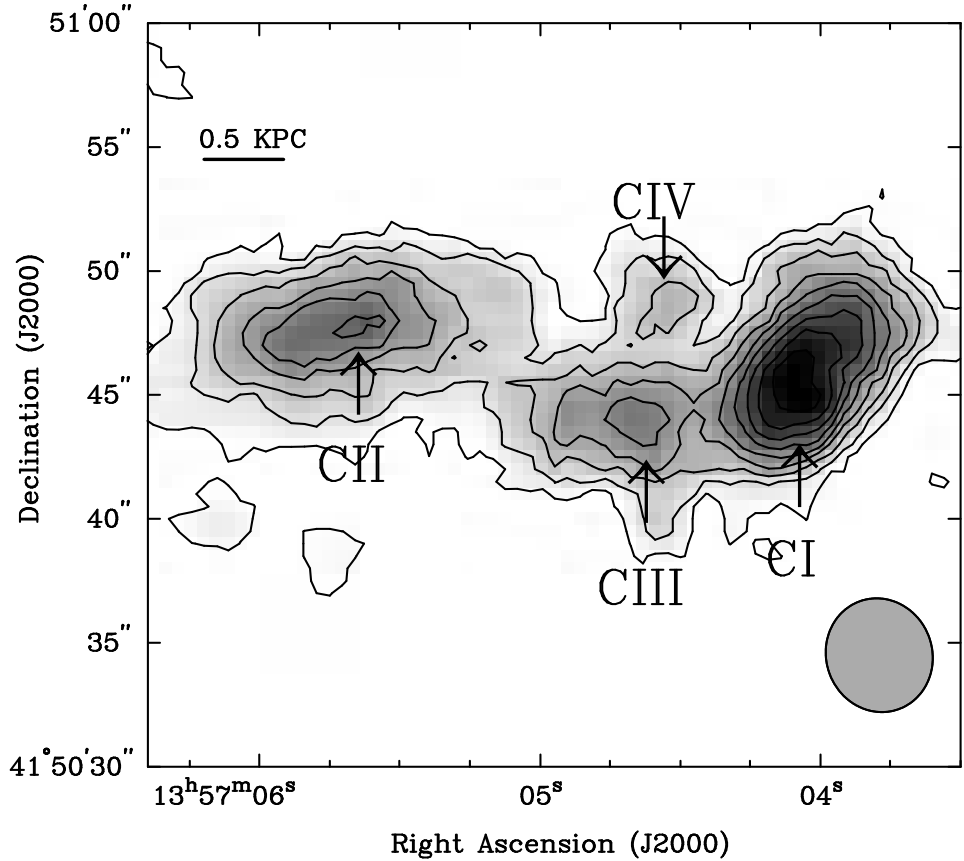,angle=-90,width=5in}}
\caption[CO Map]
{Nine-element BIMA CO (J=1-0) total intensity map, with a naturally
  weighted beam of 4$\farcs$62 $\times$ 4$\farcs$23 and contours at
  2,4,6,8,10,12,14,16, and 18 $\sigma$(1$\sigma$ = 1.85Jy~\kms~\bm ).  CI
  and CII identify the ``twin peaks'' which coincide with the inner end of
  the dust lanes.  CIII marks the third peak, located interior to the
  twin peaks. CIV is coincident with the brightest star forming
  region in the nucleus\label{cofig}}
\end{figure}

We observed CO (J=1-0) line emission in NGC~5383 with the BIMA
(Berkeley-Illinois-Maryland Association)\footnote{The BIMA Array is
  partially funded by a grant from the National Science Foundation} 
 array in three different
configurations of six dishes from February-June 1994, and in a 
compact configuration of nine dishes from July-August 1996. 
The phase and pointing center for these observations was
$\alpha$(J2000) = 13$^h$57$^m$04$^s$.54, $\delta$(J2000) =
41$^o$50$^\prime$ 46$\farcs$00.  
The correlator window was centered at V$_{LSR}$ = 2245 \kms~with a
total bandwidth of 424 \kms~ and a resolution of 4.1 \kms.
In each case the line emission was observed in the upper sideband.
The projected baselines ranged from 2.2 to 87.5 k$\lambda$ and the
single-sideband system temperatures ranged from 500 to 1100 K.  
We flagged the data with anomalous visibilities resulting from
shadowing of an antenna at low elevation and spurious electronic
spikes (birdies and correlator window edges).  
The complex instrumental gain was calibrated using the quasars
1153+495, 1419+543 or 1310+323, observed every 20-30 minutes. 
The absolute flux calibration was determined from observations of
Uranus or Mars and the structure of the IF band pass was determined
from observations of the quasars 3c273 or 3c279.   
We imaged the calibrated data set into 10 \kms~channels using natural
weighting to achieve maximum signal to noise ratio and then cleaned
the dirty maps using the Hogb\"om algorithm
\markcite{hogbom74}(Hogb\"om 1974) .
Next we employed an iterative phase-only self-calibration process
\markcite{thompson86}\markcite{regan95}(Thompson, Moran, \& Swenson
1986; Regan et al. 1995) which used the CLEAN components as model 
inputs for the iterations.  
The self-calibration process did not significantly improve the results
because our initial atmospheric phase calibration was good.  
In other words, our data are limited by thermal noise rather than the dynamic
range.  
The synthesized beam in the maps is 4$\farcs$62 $\times$ 4$\farcs$23 at a
position angle of 17 degrees.  
The final channel maps have a noise level of 50 mJy beam$^{-1}$.
We summed over all pixels in the data cube with $\mid$S$_{\nu}$$\mid$
$>$ 2$\sigma$ to form a velocity-integrated total intensity map which
is shown in Figure \ref{cofig}.  

\subsection{ CO Single-Dish Observations}

\begin{figure}[!htb]
\centerline{\psfig{figure=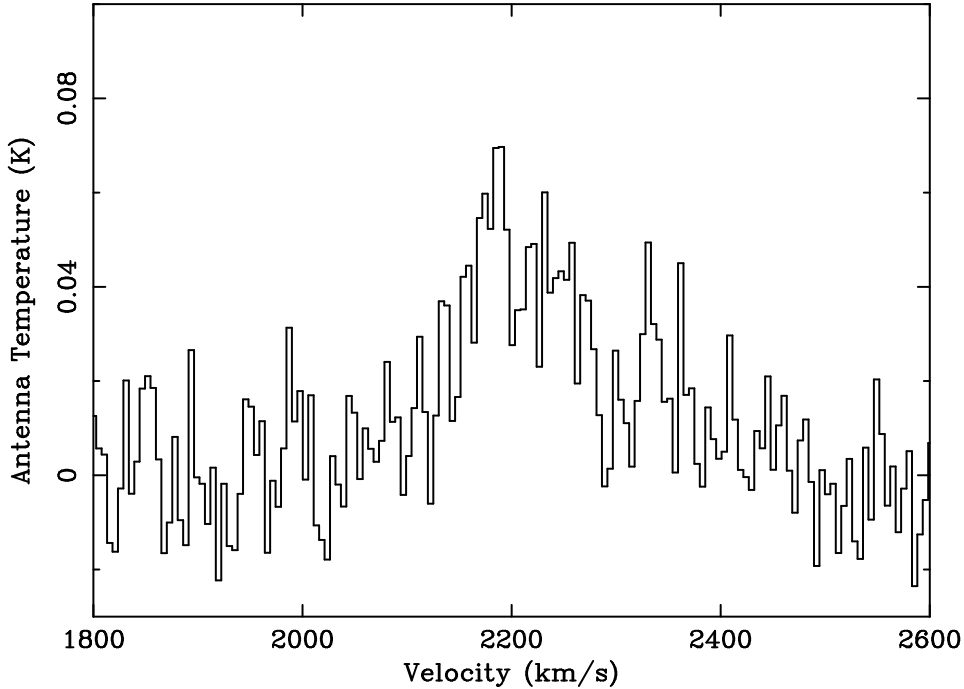,angle=-90,width=5in}}
\caption[CO Single Dish Spectrum]
{Single dish spectrum of NGC~5383 obtained with the NRAO 12m single
  dish telescope.  Integrating from 2060 to 2430 \kms, we get a total
  flux of 9 K km/s.  \label{cosdfig}}
\end{figure}

We obtained a spectrum of the center of NGC~5383 in the CO (J=1--0)
emission line  with the single dish, 12m NRAO\footnote{The National
  Radio Astronomy Observatory is a facility of the National Science
  Foundation, operated under cooperative agreement by Associated
  Universities, Inc. telescope at Kitt Peak} telescope on 01 December
1998.   
The half-power primary beamwidth of the telescope is 53$\arcsec$ and the
pointing accuracy is $\pm$ 5$\arcsec$.  
We used the spectral line beam position switching observing mode with
a 4$\arcmin$ throw for the off position.  
The total on-source integration time was nine minutes with a 
system temperature of $\sim$350 K.  The filter banks were configured
in the 2 $\times$ 2 MHz series mode for a total bandwidth of 512 MHz.
An unstable oscillation in one channel prevented us from using data from both
channels.  The measured rms noise in a 2 MHz channel is 19.5 mK.   A
zeroth order baseline was subtracted from each of the three scans
before they were co-added; the resulting spectrum is shown in Figure
\ref{cosdfig}.  

\section{RESULTS}

\subsection{CO Morphology} \label{comorph}

CO emission is detected in the central 3.5 kpc diameter region of the NGC
5383.  
We find two prominent CO peaks (labeled CI, CII in Figure \ref{cofig})
in the total intensity map, corresponding to the inner termini of the
dust lanes\footnote{In this paper, we use the term dust
  lane and bar dust lane interchangeably.  All references to dust
  lanes always refer to the main bar dust lanes}. 
In addition to these two peaks, we see a third peak (labeled CIII in
Figure \ref{cofig}) east-southeast of peak I.
Peak CIII lies along the northwestern dust lane after it has curved
and is heading east across the nuclear region.  
There is CO emission in an even weaker feature (labeled IV) 5$\arcsec$
north of CIII.  This feature coincides with the most intense 
star forming region in the nucleus.

Within the primary beam (53$\arcsec$ FWHM) of the NRAO 12m telescope,
the total CO flux in the BIMA interferometer maps is $\sim$175 Jy
\kms.  This is about 60\% of the flux (9 K \kms) measured from the
NRAO spectrum, assuming a conversion factor of 33 Jy K$^{-1}$ for the 12m
telescope.  Most of the CO detected by the interferometer is
concentrated in the central 35$\arcsec$ of NGC 5383.  The CO flux in
this 35$\arcsec$ region (165 Jy \kms) can be converted to a total
H$_2$ mass using the standard equation M(H$_2$) = 1.1 $\times$ 10$^4$
D$^2$ S$_{CO}$ (\markcite{kenney92}Kenney et al. 1992), where D is in
Mpc and S$_{CO}$ is in Jy \kms.  For the adopted distance of 32.4 Mpc,
the total molecular gas mass is $\sim$2 $\times$10$^9$ \Msun.

\subsection{Dust Extinction Morphology} \label{dustmorph}

\begin{figure}[!htb]
\centerline{\psfig{figure=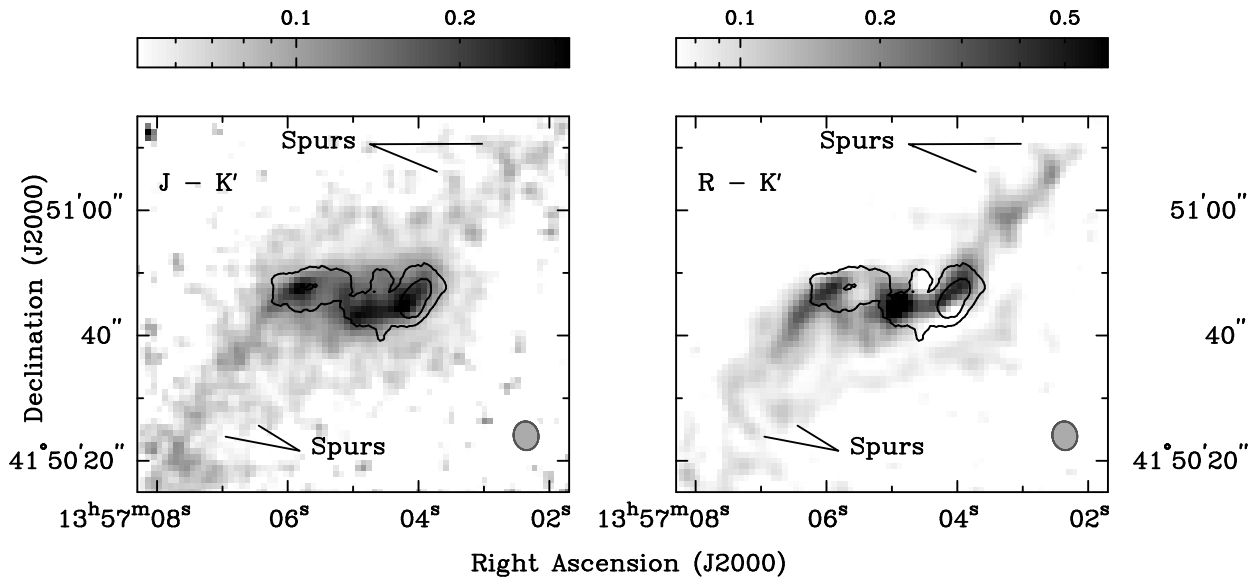,angle=-90,width=6.5in}}
\caption[Dust extinction in NGC~5383]
{Dust extinction in NGC~5383 as traced by optical-infrared and
  infrared-infrared colors.  The resolution of both maps is
  2$\farcs$4.  Left panel: J$-$K$^\prime$ color map
  (grey scale) of NGC~5383 tracing regions of high dust extinction.
  The wedge above indicates the color excess in magnitudes.  Contours
  (4,8,12, and 16 $\sigma$) of CO (J=1-0) are overlaid to indicate the
  correlation between CO emission and dust extinction. The
  southeastern dust lane is clearly seen all the way to the bar end.
  The northwestern dust lane is harder to distinguish but its general
  structure and spurs similar to the R$-$K$^\prime$ map can be seen.
  Right panel: R$-$K$^\prime$ color map tracing regions of low dust
  extinction with the same CO contours as the left panel. The
  locations of the dust spurs are indicated in both panels.
  \label{dustfig}} 
\end{figure}

We have divided the R\footnote{We do not present the broad-band images
  here but we refer the reader to the paper by 
  Burbidge et al. (1962) and the Carnegie Atlas
  (\markcite{sandage94}Sandage \& Bedke, 1994) for optical images of
  NGC~5383.  In these images, note the dust lane structure which we
  highlight using color maps.  Also note that the spurs which are
  important in our discussion of bar star formation may be easier
  to see in these images}
and J band images into the K$^\prime$ band image to form color
  maps (see Figure \ref{dustfig}).  

Prior to division, the images were registered against each other and
the higher resolution images were smoothed to match the lower
  resolution (2$\farcs$4) of the K$^\prime$ band image.  
In both maps the darker grey scale corresponds to redder colors
(higher extinction).  
Although it is difficult to estimate the dust content in galaxies, it
has been shown that optical-infrared colors such as R$-$K$^\prime$ and
infrared-infrared colors such as J$-$K$^\prime$ can be used to
estimate the amount of dust using multiple scattering radiative
transfer models \markcite{regan95}(Regan et al. 1995);
however, such models require images at multiple optical and infrared
wavelengths to confidently parameterize the optical depth and the
ratio of dust scale height relative to the stars.   
Even though the lack of multiple images at different wavelengths
prevents us from this quantitative analysis, we can still gain a
qualitative understanding of the dust distribution from the two colors
available to us.  

In both maps, the main bar dust lanes show up as regions of high
extinction.  
In the R$-$K$^\prime$ map, the northwestern dust lane is relatively
straight and 
narrow with at least two spurs (indicated by arrows in Figure
\ref{dustfig}), which merge into the dust lane, and end in areas with
high levels of dust extinction.  
In the southeastern dust lane the spurs and the dust lane are
difficult to distinguish from each other, and the dust lane
characteristics are harder to describe.   
However, this dust lane is seen more clearly in the J$-$K$^\prime$
color map, where its appearance and structure is similar to the
northwestern lane.  
Note that the northwestern dust lane, on the other hand, is not clearly visible
in the J$-$K$^\prime$ map.  
The differences between the two color maps could be due to a variety
of reasons such as different dust lane thicknesses, geometry of
the dust/star mixture (\markcite{witt92}Witt 1992), systematic
reddening of the near side of a galaxy based on its orientation in the
sky (e.g. \markcite{elmegreen99}Elmegreen \& Block, 1999), or the more 
remote possibility that the stellar populations are different between
the two sides of the bar.  

At the inner terminus of the dust lanes where the dust lanes begin to
curve, two red peaks mark locations of the highest dust extinction.  
These dust extinction ``twin peaks'' match the peaks seen in CO
emission.  
Interior to these peaks, a S-shaped curve is traced in dust
extinction similar to the curve seen in CO emission.  
At the center of the S-shaped curve we find a peak with the reddest
color in the map.  
Since this central peak coincides with the peak of K$^\prime$  band
emission it probably reflects excess K$^\prime$ band emission from either a
change in the underlying stellar population (excess red giant stars), or hot 
dust (e.g., \markcite{thatte97}Thatte et al. 1997;
\markcite{marco98}Marco \& Alloin 1998), instead of a large amount of
dust extinction. 
The S-shaped structure, the dust lanes and the spurs, however, cannot
be due to K$^\prime$  band emission excesses because the
K$^\prime$  band emission is much smoother (Figure \ref{kbandfig});
hence, these must be due to dust extinction.  

\subsection{\Ha Morphology} \label{halpha}

Strong \Ha line emission (Figure \ref{hafig}) is detected in the
nucleus, the bar ends and the spiral arms.  
In the nucleus the peak \Ha emission (indicated by the label HP in
Figure \ref{hafig}) is located 3$\farcs$1 northwest of the
radio-continuum peak\footnote{Radio continuum peak position adopted
  from \markcite{becker95}Becker, White, \& Helfand (1995)}.   
We can clearly trace the northwestern dust lane in absorption against
the \Ha emission as it curves inwards, 3$\arcsec$ south of HP.  
We also see a bright elongated HII region, 5$\arcsec$ southwest of
HP, along the southern edge of the dust lane.  
Similarly, there is another bright HII region 11$\arcsec$ east of HP,
near the inner terminus of the southeastern dust lane.  

HII regions are also observed at both bar ends and along the
spiral arms near the bar ends.  
Since it is unclear whether the observed HII regions belong to the
bar end or the spiral arm, we use the term ``bar end / spiral arm HII regions''
to describe any HII regions in this area.  
The bar end / spiral arm HII regions mostly trail the bar, although
there is at least one bright HII region on the leading side of the bar.  
The HII regions extend for about 4 kpc on the trailing side of the
spiral arms and about half that distance along the leading side.  
The study of star formation activity in these regions
is interesting because of the changing environment between the bar
ends and spiral arms (\markcite{kenney91}Kenney \& Lord 1991).
However, in this paper, we limit the scope of our study and 
do not study the star formation activity in these regions.  

Instead, we choose to focus on the star formation activity in the
bar.  While the nuclear region and the bar ends have strong \Ha emission,
the bar itself is comparatively weak; this pattern is consistent with
observations of other early-type 
barred galaxies \markcite{garcia96} \markcite{koopmann96} 
\markcite{kenney96}\markcite{phillips93}(Garcia-Barreto et al. 1996;
Koopmann \& Kenney 1996; Phillips 1993).  
Most of this limited star formation activity in the bar of NGC~5383 is
found in one or two bright star formation sites on each side of the
bar, along its leading edges.  
These \Ha peaks are best seen in the contour image of \Ha emission
in Figure \ref{hadust} and their significance is discussed in \S
\ref{stardust}.

\section{DISCUSSION \& ANALYSIS}

\subsection{Why Choose Hydrodynamic Models?} \label{whyhydro}

Modeling studies of gas flows in bars constitute an extensive
subfield and we refer the reader to excellent reviews in
\markcite{sellwood93}Sellwood \& Wilkinson (1993),
\markcite{a92a}\markcite{a92b} Athanassoula (1992a,b),
\markcite{teuben96}Teuben (1996) and references therein.  
One can broadly categorize these modeling
efforts into two main categories:  a) cloud-based/sticky particle simulations
which treat the gas as a set of distinct particles which react to
collisions based on some prescription for cloud collisions, and b)
hydrodynamic (grid-based or smooth particle hydrodynamics (SPH))
simulations which treat the 
gas as an ideal fluid obeying the basic hydrodynamic equations.
Both classes of models have advantages and
disadvantages and both can provide adequate frameworks in which
observations can be interpreted.  

Both classes of models produce dust lanes but the hydrodynamic models
alone can produce the straight dust lanes seen in strongly barred
galaxies such as NGC~5383.  
Both classes predict gas inflow of approximately the same order of
magnitude but the exact route followed by the gas is
very different in each class of models.  
In the hydrodynamic model, the gas undergoes a shock at the dust lane
and is forced to flow directly inwards in the dust lane
(\markcite{regan97}RVT97, \markcite{a92b}Athanassoula 1992b).  In the
cloud-based models, the gas clouds in the dust lane behave like those
found in galaxy spiral arms.  The dust lane is formed because the
clouds crowd together for a longer time.  However, unlike the hydrodynamic
models, they eventually cross and leave the dust lane on the leading
side of the bar.  
In this model, the clouds can collide, lose
angular momentum and spiral down to the center in a few orbits
(\markcite{combes85}Combes \& Gerin 1985).  

For strong bars, the hydrodynamic model gas flow is the best
match to the observed velocity fields (e.g., NGC~1530
\markcite{regan97}RVT97).  
It is also particularly successful at producing the high shear and shock seen
across the bar dust lanes (\markcite{a92b}Athanassoula 1992); these
cannot be produced by the cloud-based/sticky particle models
(\markcite{regan99}Regan, Sheth \& Vogel 1999, hereafter RSV99).  
Moreover, a recent study of gas kinematics in seven barred spirals
also shows that the observed gas kinematics are consistent with the
hydrodynamic models and not with cloud-based simulations
(\markcite{regan99b}RSV 1999).  
So even though the molecular ISM is certainly not a diffuse, ideal
fluid, it appears that the hydrodynamic models are better at
reproducing observations than
cloud-based/sticky particle models, especially for describing
observations of gas kinematics and morphology in strong bars.  
Hence we choose to compare the CO and dust extinction morphology of
NGC~5383 to a hydrodynamic model; the model chosen is the
\markcite{piner95}PST95 standard model. 

\subsection{Comparing Hydrodynamic Model Predictions to
  Observations}\label{hydromorph}  

\begin{figure}[htbp]
\psfig{figure=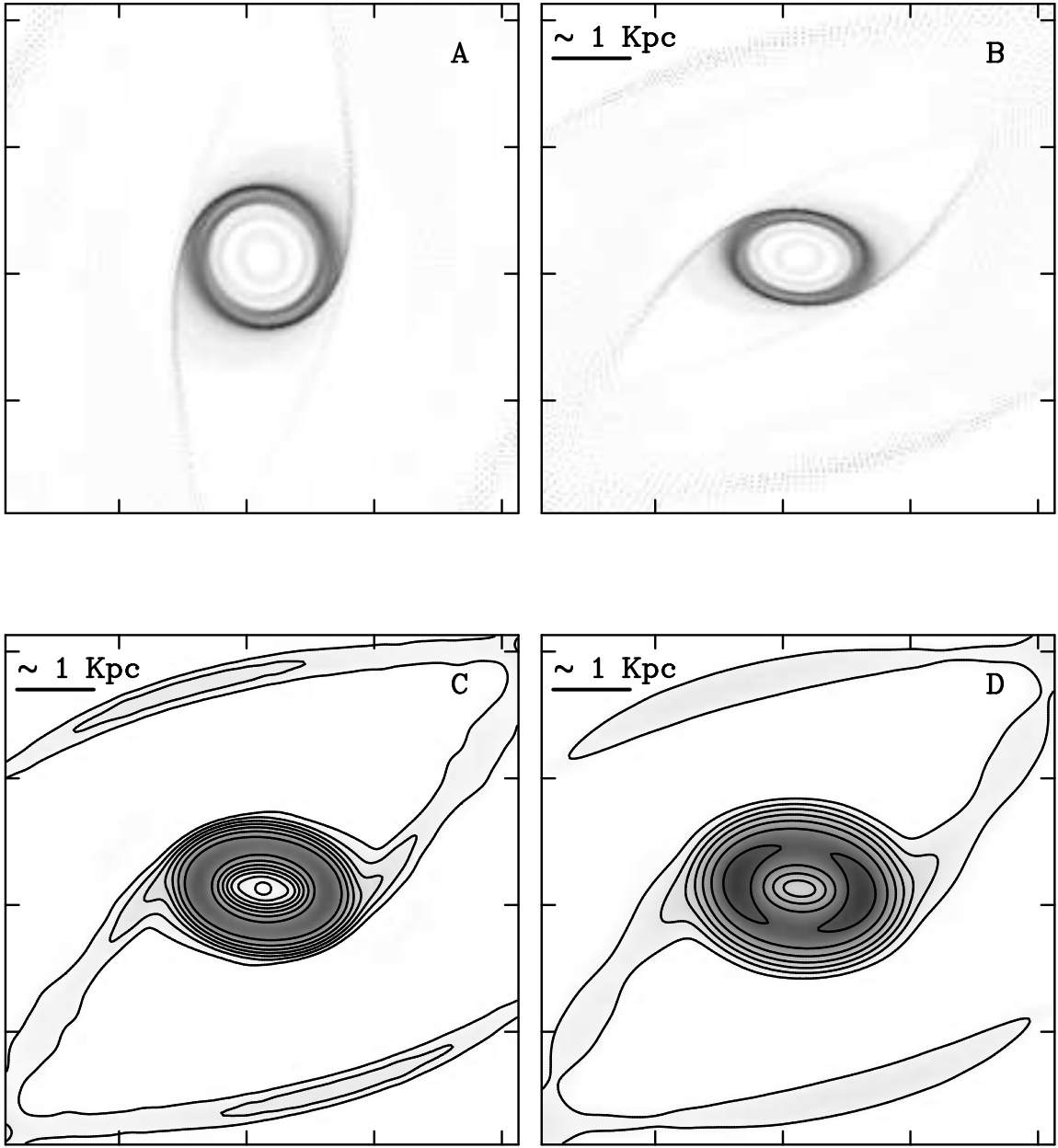,width=6in}
\caption[Model gas density]
{Top left: a) The PST95 hydrodynamic model.  Top right: b)
  The model projected to match the orientation of NGC~5383.  Bottom
  left: c) The projected model in panel (b) is smoothed to match the
  resolution of the dust extinction maps.  Bottom right: d) The
  projected model smoothed to match the resolution of the CO
  maps. Note the increasing density  inwards in the dust lane.  The
  gas on the outside of the dust lanes (the ellipsoidal 
  feature) connects to the bar end and remains to be studied.  The
  contours in the smoothed models are spaced at 2, 4, 8, 16, 32, 64,
  128, 256 $\times$ an arbitrary density.\label{modmorph}}
\end{figure}

In order to compare the CO emission and dust extinction distribution
to the hydrodynamic model, we projected the PST95 model 
(Figure \ref{modmorph}a) to match the orientation of NGC~5383 (Figure
\ref{modmorph}b) and smoothed the model to match the resolutions of
the dust extinction and CO emission maps (Figures \ref{modmorph}c,d
respectively).

The model gas distribution shows an extremely dense nuclear ring with
two peaks along the ring near the bar major axis, consistent with
PST95.  However, we emphasize that the location of these peaks varies in
the simulations.  Over time additional peaks often appear and disappear
in the ring, downstream of the dust lane terminus.  
The density contrast between the peaks and the dust lane is large; in
the unsmoothed model, the peaks have a column density which is $\sim$
25 to 120 times higher than that in the dust lane, and the average
ring column density is $\sim$ 25 to 60 times larger.
Smoothing the models reduces the average density contrast by
$\sim$50\% at the resolution of the dust extinction maps and by
$\sim$60\% at the resolution of the CO emission maps.  
We also find that the gas density in the dust lane increases inwards
along the dust lanes; from the outer terminus to the inner terminus
the density increases by a factor of $\sim$4.  
The model also produces offset bar dust lanes and an outer, low
density ellipsoidal feature.  

When we compare Figure \ref{modmorph} to Figures \ref{cofig} \&
\ref{dustfig}, we find significant differences between the model gas
morphology and the observations.  
The most striking disparity is the density contrast between the dust lane
gas and the ring/peak gas.  
The contrast in the dust extinction maps between the dust lane and the
twin peaks is 1.3 in the J$-$K$^\prime$ map and 1.5 in the
R$-$K$^\prime$ map\footnote{These values are expressed as flux ratios
  and not magnitudes}.    
At the same resolution, the model column density contrast is an order of
magnitude ($\sim$ $\times$8$-$40) larger.  
Since CO is not detected in the dust lanes, only a lower limit to the
CO brightness ratio between the peaks and the dust lane can be 
calculated. In NGC~5383, this ratio is $\sim$20, similar to that
observed in NGC~3351 and NGC~6951 (\markcite{kenney92}Kenney et
al. 1992) \footnote{We assumed that the lowest contour in Figure 1 of
  \markcite{kenney92}Kenney et al. (1992) is at 2$\sigma$}.  
In other galaxies where CO is detected in the dust lanes (e.g. NGC
1530, see Figure 11 or Table 3 in \markcite{downes96}Downes et
al. 1996, or NGC~7479, see Figure 5 in \markcite{quillen95} Quillen et
al. 1995), the CO brightness ratio between the ring/peak and the dust lane
gas is only $\sim$5$-$10; at the same resolution, the model density
contrast is $\sim$15$-$70.  
One reason for the high CO brightness ratio in NGC~5383 may be that
the CO emissivity per H$_2$ molecule is enhanced in nuclear region
(\markcite{regan95}Regan et al. 1995) or at least at the twin peaks.  
Since variations in CO emissivity may determine the CO brightness
ratio, dust extinction is perhaps a better tracer of the gas column density
because the gas to dust ratio is better constrained than the CO to
H$_2$ conversion factor (\markcite{sodroski95}Sodroski et al. 1995;
\markcite{regan00}Regan 2000) \footnote{This would mean that the total
  molecular gas mass calculated from the standard CO to H$_2$
  conversion factor is an upper limit}.
Therefore comparing the dust extinction maps to the model, we
conclude that the ratio of the ring to the dust lane gas surface
density is an order of magnitude larger than seen in the observations.  

Another striking difference between the model and the observations is
the central morphology of gas and dust.  
In the PST95 study, a nuclear ring always formed whenever
the bar had offset dust lanes (i.e. had an ILR), and whenever the
bar was thick (axial ratio $<$ 5).  
In NGC~5383 both of these conditions are met, yet there is no sign of a
nuclear ring; instead, as described in \S \ref{comorph} and \S
\ref{dustmorph}, the gas and the dust are distributed in a S-shaped
pattern with three peaks.  
In some galaxies, e.g. NGC~1530 (\markcite{regan95}Regan et al. 1995),
higher resolution CO maps have resolved the twin peaks structure into a
broken ring or partial spiral arms.   
In NGC~5383, however, higher resolution uniform weighted maps (not
shown) do not reveal a ring; they show the same triple peak structure
seen in the lower resolution CO map.  
Also, the dust extinction maps, which are at a higher resolution than
the CO maps, do not show a ring.  
In some other ``twin peak'' galaxies, e.g. NGC~6951, NGC~3351 and
M101, a nuclear ring of HII regions has been observed
(\markcite{kenney92}Kenney et al. 1992).  
However, we see no evidence of an \Ha ring in
our data; the \Ha emission is amorphous and is dominated by the bright
\Ha peak north of the nucleus (see Figure \ref{hafig}).  
While it is possible that an \Ha ring can be obscured by highly
variable and patchy dust extinction in the nuclear
region (\markcite{phillips93}Phillips 1993), the dust distribution
in NGC~5383 is not ring-shaped and cannot hide an HII ring.  

At this point, it worthwhile to mention that the CO morphology in the
center of bars can have a rich variety of morphologies in addition to
``twin peaks.''  As \markcite{kenney96}Kenney (1996) points out, there
are spiral arms as in NGC~6951 \markcite{kenney96}(Kenney 1996),
filled exponential disks as in NGC~3504 (\markcite{kenney93}Kenney, Carlstrom
\& Young 1993) or NGC~4102, and rings or partial rings as in NGC~4314
\markcite{benedict96}(Benedict, Smith \& Kenney 1996).  
So although we are only examining a single case, the failure of the
hydrodynamic models to produce these morphologies in general is
problematic.  

Lastly, the models differ from the NGC~5383 observations because they
produce a low density, ellipsoidal feature which encircles the bar region.  
The ellipsoid is present in the models throughout the simulation, and
while its density varies, the density is always fairly low.   
This feature is not detected in our observations, and we
are not aware of any barred galaxy with such a molecular gas feature;
this may be due to the lack of sensitivity of current millimeter
wavelength interferometers.   
The ellipsoidal feature probably correspond to the
inner ring seen in optical images of other galaxies presented in the
review by \markcite{buta96b}Buta \& Combes (1996).    
The inner ring is usually a ring or ellipse of star formation which
encircles the bar and may be associated with a 4:1 ultraharmonic resonance
(\markcite{buta96b}Buta \& Combes 1996).   
Good examples of this feature are seen in the
continuum-subtracted \Ha images of NGC~3504 (\markcite{kenney93}Kenney et
al. 1993) and NGC~6782, UGC~12646, IC~5240
(\markcite{crocker96}Crocker, Baugus \& Buta 1996). 
Before we can discuss the importance and relevance of this feature to
gas flows in bars, we would need to obtain additional observations of
molecular gas emission and dust extinction to identify the feature,
and we would need to study the models more carefully to establish the
role of the gas in this ellipsoidal feature.  

While there are many discrepancies between the model gas distribution
and the observations, there are also some similarities.
As already shown by \markcite{a92b}A92, both straight and curved bar
dust lanes are reproduced as a function of the bar strength by the
hydrodynamic model.   
In our study, the model (Figure \ref{modmorph}) predicts increasing
gas density inwards in the dust lane; this matches the observations of
CO emission and dust extinction (Figure \ref{cofig}, \ref{dustfig}).  
A similar trend is also seen in the barred spiral NGC~1530 (See Figure
2.10 in \markcite{regan95} Regan et al. 1995, or Figure 11 in
\markcite{downes96} Downes et al 1996).  

In summary, the hydrodynamic model reproduces the observed offset dust
lanes and the observed increase in gas density inwards along the dust
lane.  
But there are key differences between the model and the observations.
The model predicts a nuclear ring morphology with an extremely high
density contrast between the ring and the dust lanes, whereas
observations show a S-shaped morphology with three peaks with a
relatively low density contrast between the dust lanes and the peaks.

\subsection{Understanding the Differences between the Model and the
  Observations}\label{understand} 

\subsubsection{Is a Nuclear Bar Affecting the Observed Gas Density and
  Morphology?}

\begin{figure}[!htb]
\centerline{\psfig{figure=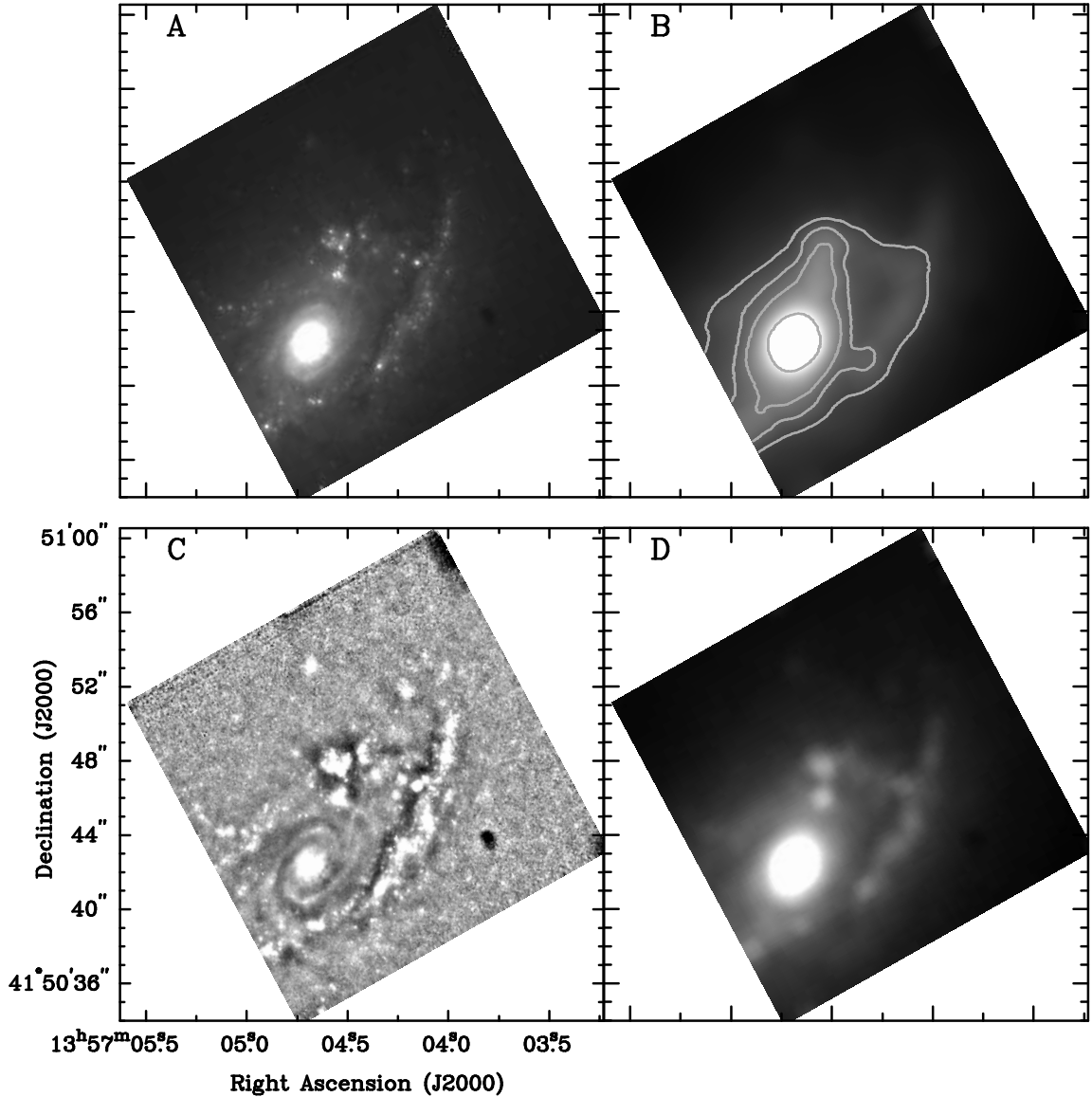,width=6in}}
\caption[NICMOS Snapshot of the center of NGC~5383]
{Top left: a) Original NICMOS image showing the nucleus and the northwestern
  dust lane in absorption.  Top right: b) Same image as in panel
  (a) smoothed by a Gaussian with a FWHM of 14 pixels.  Arbitrary
  contours are drawn on top of the image to show the fake ``nuclear bar.''
  Bottom left: c) Unsharp masked image created by subtracting panel
  (b) from panel (a). The trailing spiral pattern is
  clearly seen all the way into an unresolved core.  Bottom right: d)
  Same image as in panel (a) smoothed by a Gaussian with a FWHM of 7
  pixels.  Notice the fake ``ring'' around the nuclear bar.
  \label{nicfig}} 
\end{figure}

One reason for the absence of the predicted morphology could
be the presence of a nuclear bar which could drive gas inwards from
the twin peaks.  
We searched for a nuclear bar in a high resolution, high dynamic range
NICMOS image (Figure \ref{nicfig}a).  

The image shows a relatively smooth nucleus with a collection of star
forming regions north northwest of the nucleus.  These correspond to
the peak HP seen in the lower resolution \Ha image (Figure
\ref{hafig}).  The main bar dust lane can also be seen in absorption 
as it travels inwards towards the nucleus.  
There is no evidence of a nuclear bar at the center of NGC~5383.  

Since the nucleus in NGC~5383 is quite bright, it is possible that
weaker structure in this region may be difficult to discern.  To
investigate this possibility we applied the technique of unsharp
masking where we subtracted a 14 pixel FWHM Gaussian smoothed image
(Figure \ref{nicfig}b) from the original image (Figure \ref{nicfig}a).
Pioneered by \markcite{malin79}Malin \& Zealey (1979), this technique
allows one to distinguish weaker features from brighter ones by
reducing the contribution of the bright background; however, we
caution that the unsharp masking technique is not perfect and detailed
studies of nuclear structure may benefit from multiple techniques.
For our purposes the unsharp masking technique is adequate; the
resulting image is shown in Figure \ref{nicfig}.  It shows a
well-defined two arm, trailing spiral pattern which starts from the
main northwestern dust lane and winds its way down to an unresolved
core; there is no evidence of a nuclear bar.  The observed trailing
pattern is not unique to NGC~5383; central regions of other galaxies
also show similar nuclear trailing spirals (e.g. NGC~3982, NGC~3032,
Mkn 573 \markcite{regan99a}Regan \& Mulchaey 1999 and references
therein).  This trailing spiral arm pattern may be a mechanism for
transporting gas further inwards from the location of the inner
Lindblad resonance (\markcite{regan99a}Regan \& Mulchaey 1999) where
the gas inflow is known to stall in modeling studies.

Models have not predicted this trailing spiral pattern.  
Analytic solutions predict a leading spiral structure inside the inner
inner Lindblad resonance (IILR) (\markcite{combes96}Combes 1996;
\markcite{yuan97}Yuan \& Kuo 1997).  
Although a trailing spiral pattern can be excited at the outer inner
Lindblad resonance (OILR), only a leading spiral is excited by the IILR.  
It is unclear whether the OILR excited trailing spiral pattern can
propogate inside the IILR.  
It is also unclear how the two spiral patterns interact and whether
the relative amplitudes of the two patterns can appropriately account for
these observations.  These issues are under investigation
(\markcite{yuan99}Yuan et al. 1999).  

The smoothed image used for the unsharp masking (lower right hand
corner in Figure \ref{nicfig}) has a resolution and quality comparable
to good ground-based images.  
In this image, we clearly see a nuclear bar-like feature at
the center of NGC~5383 even though there is no bar in the original
NICMOS image and in the unsharped masked image.  
The bar-like feature is created by the smoothing of the light from the
bright star forming regions north and south of the nucleus which leads
to the roughly north-south elongation, i.e. the fake bar.   
Smoothing the original image with a Gaussian with FWHM of 7 pixels,
one can even create a partial circumnuclear ring (see Figure
\ref{nicfig}d).  Although this is more 
difficult to see than the bar, an oval encircling the fake nuclear bar is
clearly visible in this image.  
In reality, these structures are simply artifacts of the various
smoothing techniques applied to the original image.  

This result shows that study of the nuclear structures in a galaxy
requires high resolution observations, and that ground-based images
can be inadequate and worse, misleading in the identification of these
structures.  Another important point of note is that even in the
H-band, where the effects of dust extinction and star formation are
supposed to be minimized, these NICMOS images clearly show star
forming knots and dust features.  The smoothing experiments show how
these contaminants can create false features.  So although
near-infrared observations are desirable compared to optical
observations (which are even more sensitive to these contaminants),
the near-IR data also need to have high resolution to properly
investigate nuclear features.

Thus, we conclude from this study of the NICMOS image of the center of
NGC~5383 that a nuclear bar is not responsible for the
differences between the observations and the model predictions.  

\subsubsection{Can the Hydrodynamic Model be Tuned to Account for the
  Morphology Discrepancies?} 

We considered the possibility that the hydrodynamic models could
produce offset dust lanes without producing a nuclear ring.  
The motivation for this search came from an examination of the
bi-modal outcomes (i.e. offset dust lanes with a ring versus centered dust
lanes without a ring) of the previous studies by \markcite{a92b}A92
and \markcite{piner95}PST95.  
Since these studies took rather coarse steps in parameter space, we
considered the possibility of an intermediate scenario where offset dust
lanes could exist without a ring or with some other nuclear gas
morphology. 
We studied the parameter space by varying three of the four free 
parameters in the PST95 model (the four parameters are the bar axial ratio, the central
density, the Lagrangian radius or equivalently the pattern speed, and
the quadrupole moment or equivalently the bar mass $-$ see PST95 for
details)\footnote{We set the bar axial ratio to be 4 for our
  experiment} in such a way as 
to encompass the transition region between the two outcomes described above.  
Though we found some new and interesting results from this experiment
(see Appendix A), we did not find the intermediate scenario we were
searching for; in every case where offset dust lanes formed, we found
that a nuclear ring also formed (See Figure \ref{modfig}a,b,c) . 
Hence, we come to the conclusion that gas hydrodynamic processes alone
cannot determine the nuclear region gas morphology.

\subsubsection {Can Nuclear Star Formation Account for the Discrepancy?}

Initial parameters and the length of time over which the bar has evolved both
determine the density of the ring in the model. 
Since in the PST95 model the inflowing gas simply accretes
in the nuclear ring over time, the density contrast between the ring
and the dust lanes is destined to increase over time.  
The low density contrast in the observations of NGC~5383 might suggest
that the bar in this galaxy is fairly young.  However, similar low
gas density contrast in several other barred spirals indicates that it
is more likely that the gas is somehow being depleted from the ring.
One way of depleting the gas would be to convert the gas into stars.  
\markcite{kenney93}Kenney et al. (1993) showed that in the nuclear
region of NGC~3504, the Toomre Q parameter is close to its critical
value; they suggested that the ongoing nuclear star formation could be
explained using a simple gravitational instability picture.  
\markcite{elm94}Elmegreen (1994) also showed analytically that at
typical nuclear ring radii (r $\sim$ 1 kpc), star formation can occur via
gravitational instability if the ring gas density is roughly 100 $\times$ disk
gas density (at r $\sim$ 10 kpc), assuming that the disk gas density is
nearly critical.  
In the models, the ring density is a factor of 50-200 $\times$ initial
disk density which is sufficiently high for star formation via
gravitational collapse.  
Indeed, in the nuclear region in NGC~5383 (Figure \ref{hafig}), as in
numerous other barred spirals (\markcite{phillips93}Phillips 1993;
\markcite{garcia96}Garcia-Barreto et al. 1996), we observe vigorous,
circumnuclear, massive star formation. 

The nuclear star formation rate in NGC~5383 is at least 7 \Msun\ \yr\,  
(\S \ref{starform}).  If continuous star formation is assumed, this
rate sets a lower limit on the mass inflow rate.  
Based on measurements in other barred spirals such as NGC~7479 and
NGC~1530, which are similar to NGC~5383, it is likely that the
mass inflow rate is lower (\markcite{quillen95}Quillen et
al. 1995 and \markcite{regan97}RVT97 measure inflow rates of 1--4
\Msun\ \yr).  
But the mass inflow rate is a difficult quantity to measure and these
numbers are highly uncertain.  
Still it is unlikely that the high nuclear star formation activity can be
supported by a proportionally high mass inflow rate over an extended period
of time (for example, an inflow rate of 10 \Msun\ \yr, would deplete
the total gas reserve (10$^{10}$ \Msun) of a typical galactic disk in
only a  billion years).  

One can also compare the total gas mass in the center of NGC 5383 to
the current star formation rate.  
Using the standard CO-H$_2$ conversion factor, the total molecular gas
mass in the center of NGC 5383 is 2$\times$10$^9$ \Msun.  
Given that the dust and the CO intensities differ between the
bar and the nuclear region, and given that the dust is probably a
better tracer of the molecular gas mass, the total molecular gas mass 
calculated from the standard conversion factor is probably an upper
limit to the total gas mass in the circumnuclear region.  
Therefore at the observed star formation rate of 7 \Msun\ \yr, the gas
reservoir at the center of NGC 5383 can be depleted in less than
3$\times$10$^8$ yrs.  
So, as has been suggested before, circumnuclear star formation in NGC
5383 is probably an intense and episodic event (\markcite{shlosman92}
Shlosman 1992).  
Thus the observed low density contrast can be attributed to gas depletion
via circumnuclear star formation.  The simulations do not reproduce
the observations because, although star formation can occur at the
model densities, star formation is not modeled.  

\subsection{Star Formation Rate in NGC~5383} \label{starform}

We used the \Ha luminosity to estimate the star formation rate in
various regions in NGC~5383 (see Table \ref{sfrtab}).  
We measure an \Ha flux of 2.42 $\pm$ 0.09 $\times$ 10$^{-12}$ \ecs~ in the
central 3.5 kpc diameter region and a total \Ha flux in the galaxy of
4.37 $\pm$ 0.36 $\times$ 10$^{-12}$ \ecs.
We converted the observed \Ha flux to a star formation rate using the
empirical equation derived by \markcite{kennicutt83}Kennicutt (1983),
\begin{equation}
SFR(total) = \frac{L(H\alpha)}{1.12 \times 10^{41}}~M_{\sun}~yr^{-1} 
\end{equation}
In deriving this equation Kennicutt assumed an average intrinsic dust
extinction of 1.1 magnitudes in a galaxy.  
Since the dust extinction can vary dramatically in star forming
regions \markcite{phillips93}(Phillips 1993), the exact optical depth is
difficult to determine; hence, we use Kennicutt's value to correct the
observed fluxes for dust extinction.  
Using an adopted distance of 32.4 Mpc, we calculate the total star
formation rate in NGC~5383 to be 12.6 \Msun\ \yr, of which 7 \Msun\ \yr,
are formed in the nuclear region.  
For any single galaxy, this type of calculation of the star
formation rate is uncertain by a factor of 2
(\markcite{kennicutt83}Kennicutt 1983).  
A previous calculation using a different technique and the H$\beta$
ionization line by \markcite{duval83} Duval \& Athanassoula (1983)
resulted in a nuclear star formation rate of 8 \Msun \yr; our results
are consistent with their value.  
The star formation rate at the northeastern and southwestern bar ends
is 1.9 \& 3.2 \Msun\ \yr, respectively, smaller than the star
formation rate in the nuclear region.   

\begin{deluxetable}{lrrrr}
\small
\tablewidth{5in}
\tablecaption{Star Formation Rates in NGC~5383 \label{sfrtab}}
\tablehead{\colhead{Label}&\colhead{Name}&\colhead{Area}&\colhead{Flux}&\colhead{SFR} \\
\omit & \omit & \colhead{(arcsec$^2$)}&\colhead{(\ecs)} & \colhead
{(\Msun\ \yr)}}
\startdata
H-I & Nuclear region & 1074 & 2.42 $\pm$ 0.085 $\times$ 10$^{-12}$ & 7
\nl 
H-II & NE Bar end region & 1026 & 0.65 $\pm$ 0.07 $\times$ 10$^{-12}$
& 1.9 \nl  
H-III & SW Bar end region & 1452 & 1.09 $\pm$ 0.11 $\times$ 10$^{-12}$ &
3.2 \nl
& The galaxy & 4763 & 4.37 $\pm$ 0.36 $\times$ 10$^{-12}$ & 12.6 \nl
\enddata
\end{deluxetable}

The star formation rate of 7 \Msun\ \yr, in the central 3.5 kpc of NGC
5383 is substantially higher than the average star formation rate over
the entire disks of Sb and Sc type galaxies, which varies from 0.1 to 4
\Msun\ \yr, (Kennicutt 1983).  However, such a high star formation rate
is not unusual for centers of barred spirals, and such activity is
often classified as a starburst (e.g., \markcite{jogee99}Jogee 1999;
\markcite{contini97}Contini et al. 1997).  The total mass of H$_2$
(2 $\times$ 10$^9$ \Msun) calculated earlier is also not unusual and
it places NGC~5383 in company of galaxies such as NGC~4536 and NGC~2782
(\markcite{jogee99}Jogee 1999) which are both gas rich in the nuclear
region and hosts to high nuclear star formation activity.  

\subsection{Star Formation in the Bar}\label{stardust}

Excluding the bar ends and the nuclear regions, early type barred
spirals have very low star formation activity in the bar, even though
the gas concentration is high in the bar dust lanes
(\markcite{downes96}Downes et al. 1996).  As already shown in \S
\ref{hydromorph}, the density contrast between the dust lanes and the
center is small.  Yet the star formation activity in the bar is
limited compared to the center; several explanations have
been put forth to account for this behavior.  
\markcite{tubbs82}Tubbs (1982) suggested that molecular clouds
entering the dust lane could be dispersed due to their high velocities
relative to dust lane gas.  
\markcite{a92b}Athanassoula (1992) argued that the high shear in the
bar dust lanes prevents star formation.  
\markcite{regan97}RVT97 pointed to the large divergence in gas
streamlines prior to the dust lanes, which they argued could tear apart
molecular clouds or prevent their formation.   
Recently, \markcite{reynaud98}Reynaud \& Downes (1998) argued that a
combination of the shear and shock in the dust lanes leads to the low 
star formation rate in the bar.  
These hypotheses argue effectively for the inhibition of star formation in
the bar dust lanes.  Yet some star formation does occur in the bar.
Here we investigate precisely where and under what circumstances stars
form in the bar.  

\begin{figure}[!hbt]
\centerline{\psfig{figure=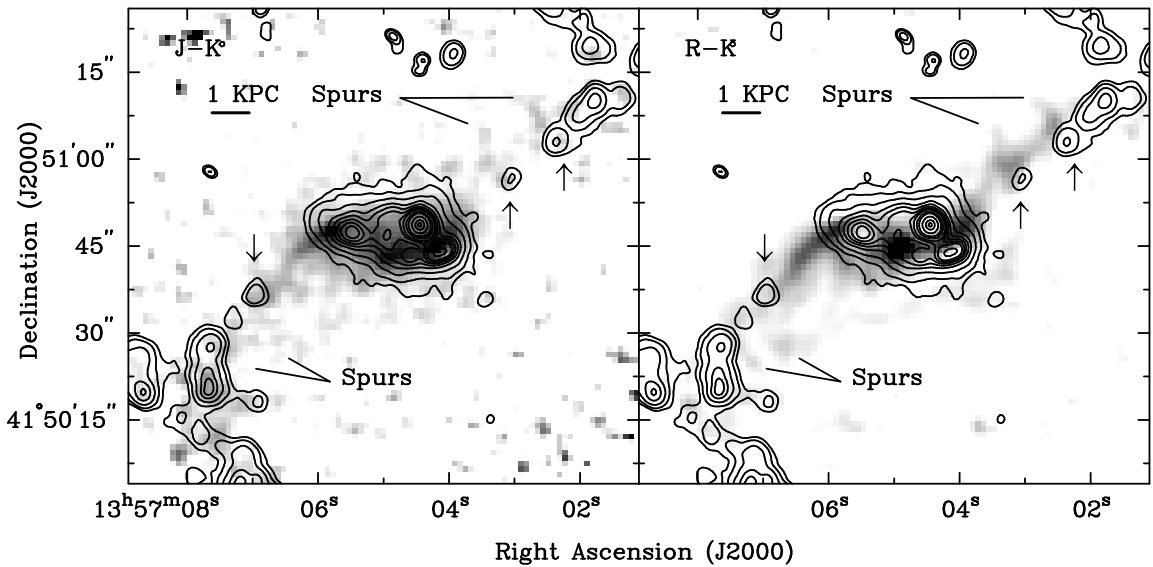,angle=-90,width=6in}}
\caption[\Ha on dust]
{A plot of \Ha contours spaced at 15, 25, 50, 100, 150, 200, 250,
  300, 400, 500, 700, 900, 1000 $\times$ 5.6 $\times$ 10$^{-18}$ \ecs,
  overlaid on the J$-$K$^\prime$ (left panel) and R$-$K$^\prime$
  (right panel) color maps.  Note the location of the \Ha peaks, shown
  by the arrows, and the dust spurs.  The \Ha peaks are all on the
  leading side of the dust lane directly across from the dust
  spurs. \label{hadust}} 
\end{figure}

The star formation sites can be traced by \Ha emission, whereas the
molecular gas can be traced by either CO emission or dust extinction.
The CO emission in the bar dust lanes of NGC~5383 is too weak to be
detected, so we assume that the dust extinction traces the molecular 
gas (as suggested observations of NGC~1530 by
\markcite{downes96}Downes et al. 1996 and observations of NGC~3627 and
NGC~2903 by \markcite{regan99b} RSV99).  
Figure \ref{hadust} shows the overlay of the \Ha emission on the dust
extinction in the bar on NGC~5383.  
Comparing the relative distribution of the star formation sites
to the molecular gas distribution in this figure, one immediately
notices two important features.  

First, the \Ha peaks are all located on the leading
side of the bar dust lane.  Second, these peaks occur 
preferentially near dust concentrations (see arrows in Figure
\ref{hadust}); these \Ha peaks are 
found at the ends of dust spurs which are most clearly seen in the
R$-$K$^\prime$ (Figure \ref{dustfig}) image.  
The \Ha peaks and the dust peaks are roughly $\sim$ 3$\arcsec$, or
$\sim$ 470 pc apart.
 
In order to place these observations in the context of the
hydrodynamic model gas kinematics, we reiterate the predictions of the
gas flow in the models.  In these models, gas flows in elliptical
streamlines until it encounters the main bar dust lane where the hydrodynamic
shock redirects the gas inwards (\markcite{97}RVT97;
\markcite{regan99}RSV99); {\it none \rm} of the
gas encountering the dust lane crosses the dust lane.  
Hence, the presence of \Ha emission on the leading side of the dust
lane is surprising.  
This is because the recent massive star formation indicated by the \Ha
emission is thought to occur only in areas of high gas
concentration such as giant molecular clouds, or in larger complexes
of such clouds called giant molecular associations
(\markcite{blitz94}Blitz 1994 and references therein;
\markcite{vogel88} Vogel, Kulkarni \& Scoville 1988); the area on the
leading side of the bar dust lane does not have high gas
concentrations.  
Therefore, we conclude that these young stars must have formed
elsewhere.  

In the bar, the only regions suitable for star formation, i.e. regions
of high gas concentrations, are the dust lanes and the dust spurs.
The dust lane, however, is thought to be an inhospitable environment
for various reasons already outlined above; the spurs are a plausible
alternative.  
Since there is a one to one correspondence between the \Ha peaks and the
dust spurs with the \Ha peaks located directly across from the spurs,
it is likely that the spurs are involved in star formation.  
In context of the hydrodynamic gas flow,  stars forming in the spurs
would continue to move along their original elliptical orbits.  These
stars would pass ballistically through the dust lane, travel to the
leading side and ionize the low density gas there.  This would be
consistent with the \Ha peaks being located directly across from the
dust spurs.  
Another important reason for the spurs to be conducive to star formation is
that, in addition to being regions of high gas density, the spurs are also
located in a region of lower shear.  
Although we do not have a full velocity field map for NGC
5383 which shows the lower shear environment, the lower shear can be
clearly seen in the gas, prior to its encounter with the dust lanes,
in the maps of other barred galaxies such as NGC~1530
(\markcite{regan97}RVT97).  

The distance between the \Ha peaks and the dust concentrations at the
end of the spurs puts a strong constraint on the velocity of the
newly formed stars.
If the stars form in the spurs, the stars must traverse $\sim$500 pc
in 10$^7$ years and therefore have velocities $\sim$ 50 \kms\ (in
the rotating frame of the bar); indeed, velocities of this
magnitude are present in the hydrodynamic model before the gas
encounters the dust lane.  
So the distance between the \Ha peaks and the dust lanes is also consistent
with star formation occurring in the spurs.

Thus we have addressed where and under what circumstances stars form
in the bar, given the predictions of the hydrodynamic model gas flow.
The observations presented here suggest the following scenario:  at
certain locations along the bar gas becomes highly concentrated and forms 
spurs.  Star formation occurs in the spurs because of the high gas
density {\it and \rm} low shear.  The newly formed stars move ballistically
through the dust lane, ionizing the low density gas on the leading
side of the bar and appearing directly across from the spurs, whereas
the gas is redirected inwards down the dust lane.  

However, one difficulty in applying this scenario to NGC~5383 is the lack of
\Ha emission in the spurs.  The spurs are simply not dense enough to
hide star formation and therefore one might expect to see \Ha emission
from ongoing star formation in the spurs. The lack of such emission in
NGC~5383 may be because we are observing very low levels star formation
activity (2-3 peaks on each side) in the bar of this galaxy.  
Indeed, observations of other galaxies do show \Ha emission on both
the leading and trailing sides of the bar dust lane (e.g., as in NGC
7479, \markcite{laine99}Laine et al. 1999);  however, it is not known
whether this emission is associated with dust spurs.  
In any case, the one-to-one correspondence between the \Ha emission
and the spurs in NGC~5383 is unlikely to be completely coincidental and
therefore it is highly likely that the spurs are involved in the star
formation activity.  

Another interesting aspect of this scenario is that it points out the
need for including gas self-gravity in the hydrodynamic model.  The
presence of the spurs clearly indicates that gas is becoming highly
concentrated along the bar.  In the region of the spurs, the divergent
gas streamlines should 
prevent cloud formation (as pointed out by \markcite{regan97}RVT97)
but the presence of the spurs shows the importance of cloud
self-gravity and cloud-cloud gravitational interactions.  These
effects are not modeled by the hydrodynamic simulations.  
So although these models are highly
successful at reproducing the velocity fields and the straight dust
lane morphology in bars, and they provide an adequate context for
interpreting the observations of \Ha emission and the spurs in the
bar, these models lack the self-consistency necessary to fully explain
the star formation process.   

\section{Conclusions} \label{conc}

The nuclear gas and dust distribution in the prototypical barred
spiral NGC~5383 cannot be reproduced by the hydrodynamic models.
Whereas the model always produces a high density contrast
circumnuclear ring, the gas and dust are observed in a low density
contrast S-shaped distribution.  We have shown that the discrepancy
cannot be eliminated by fine tuning the model.

The alternative possibility that a nuclear bar is responsible for the
differences is eliminated by using a high resolution NICMOS image.
Applying the unsharp masking technique, we find that the underlying
nuclear structure is a trailing spiral pattern.  We have also shown
how coarser resolution data, such as that found in ground--based
images, can lead to false identifications of nuclear bars and rings.

We conclude that the discrepancy between the observed and modeled gas
distribution is due to the absence of star formation in the models;
the vigorous 7 \Msun\ \yr\ circumnuclear star formation, which can
deplete the gas and lead to the observed low density contrast, is not
modeled by the hydrodynamic simulations.

Finally, we present an explanation for how stars may form in the bar
between the bar ends and the circumnuclear region.  In our scenario,
stars form in dust spurs before the gas encounters the dust lane.  The
spurs, unlike the high shear/high density dust lane, are more
conducive to star formation because they are regions of low shear and
high density.  Stars which form in the spurs travel ballistically
through the hydrodynamic shock at the dust lane, ionizing the low
density gas on the leading side of the dust lane, whereas the gas is
redirected down the dust lane.  Thus, HII regions can even be found on
the leading side of the main bar dust lanes.  

\acknowledgments We thank the referee, Jeff Kenney for his careful
reading of the original manuscript and his numerous useful suggestions
which have resulted in a far superior paper.  We would also like to
thank Lee Armus for obtaining the optical images and Eve Ostriker for
helpful discussions and suggestions which resulted in Figure
\ref{ilrs}.  We thank Jim Stone for providing the hydrodynamic model
code and for his invaluable insights in interpreting the models.  K.S
would like to thank M. Thornley, L. Looney, M. Wright, J. Morgan and
R.  Forster for their help with BIMA data acquisition and reduction.
K.S.  would also like to acknowledge useful discussions with S.
Veilleux, Y.  Fernandez, A. Koratkar, N. Volgenau and J. Crawford.
This research was supported by NSF grant AST 9314847 and AST 9613716.

\section{Appendix A}

Previous hydrodynamic models of gas flow in a barred potential have
resulted in one of two outcomes: either offset dust lanes and a
nuclear ring form or centered dust lanes without a ring form.  
In this section we describe our experiment intended to find an
intermediate case where offset dust lanes could form without the
formation of a nuclear ring?  
To that end, we have explored the parameter space using the PST95 model
near the transition region using finer steps than those used by
\markcite{a92b}A92 and \markcite{piner95}PST95.  

\begin{deluxetable}{lrrr}
\small
\tablewidth{4in}
\tablecaption{Parameters for models which produced rings\label{models}}
\tablehead{\colhead{Model \#}&\colhead{r$_l$}&\colhead{q$_m$}&\colhead{$\rho_c$} \\ 
\omit&\colhead{(kpc)}&\colhead
{(10$^{10}$\Msun\ kpc$^2$)}&\colhead{(10$^{10}$\Msun\ kpc$^{-3}$)}} 
\startdata
1& 5.85 & 4.5e4& 2.4e4 \nl
2& 5.90 & 4.5e4& 2.4e4 \nl
3& 5.95 & 4.5e4& 2.4e4 \nl
4& 6.00 & 4.5e4& 2.4e4 \nl
5& 6.25 & 4.5e4& 2.4e4 \nl
6& 6.50 & 4.5e4& 2.4e4 \nl
7& 6.75 & 4.5e4& 2.4e4 \nl
8& 7.0 & 4.5e4& 2.4e4 \nl
9& 7.5 & 4.5e4& 2.4e4 \nl
10& 6.0 & 1.5e4& 2.4e4 \nl
11& 6.0 & 2.5e4& 2.4e4 \nl
12& 6.0 & 3.5e4& 2.4e4 \nl
13& 6.0 & 4.0e4& 2.4e4 \nl
14& 6.0 & 4.7e4& 2.4e4 \nl
15& 6.0 & 4.5e4& 2.3e4 \nl
16& 6.0 & 4.5e4& 2.6e4 \nl
17& 6.0 & 4.5e4& 2.8e4 \nl
18& 6.0 & 4.5e4& 3.0e4 \nl
19& 6.0 & 4.5e4& 3.7e4 \nl
20& 6.0 & 4.5e4& 4.0e4 \nl
21& 6.0 & 4.5e4& 7.4e4 \nl

\enddata
\end{deluxetable}

\begin{figure}[!htbp]
 \psfig{figure=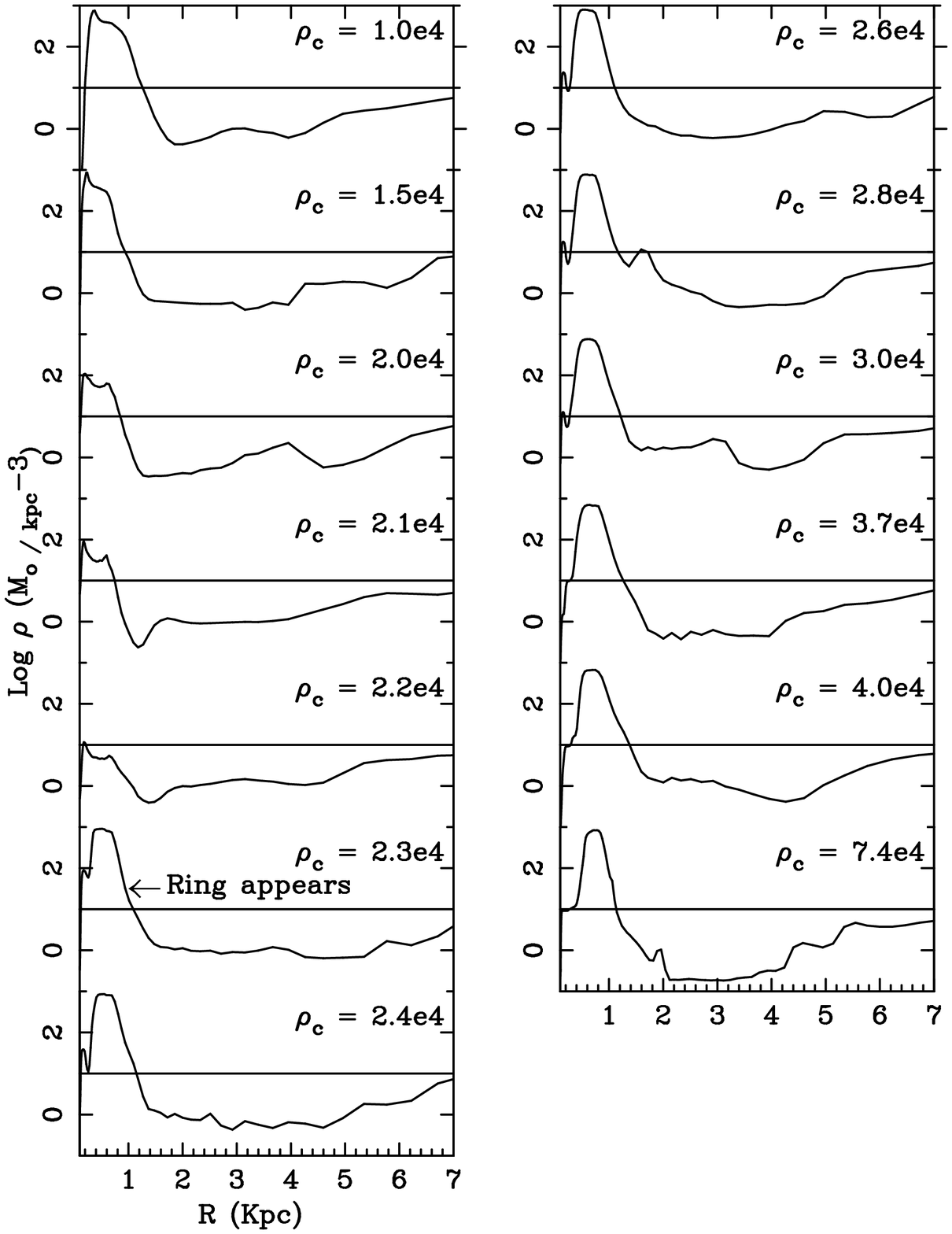,width=3.0in,angle=0,silent=1} 
 \vskip -3.9in
 \hskip 3.0in
 \psfig{figure=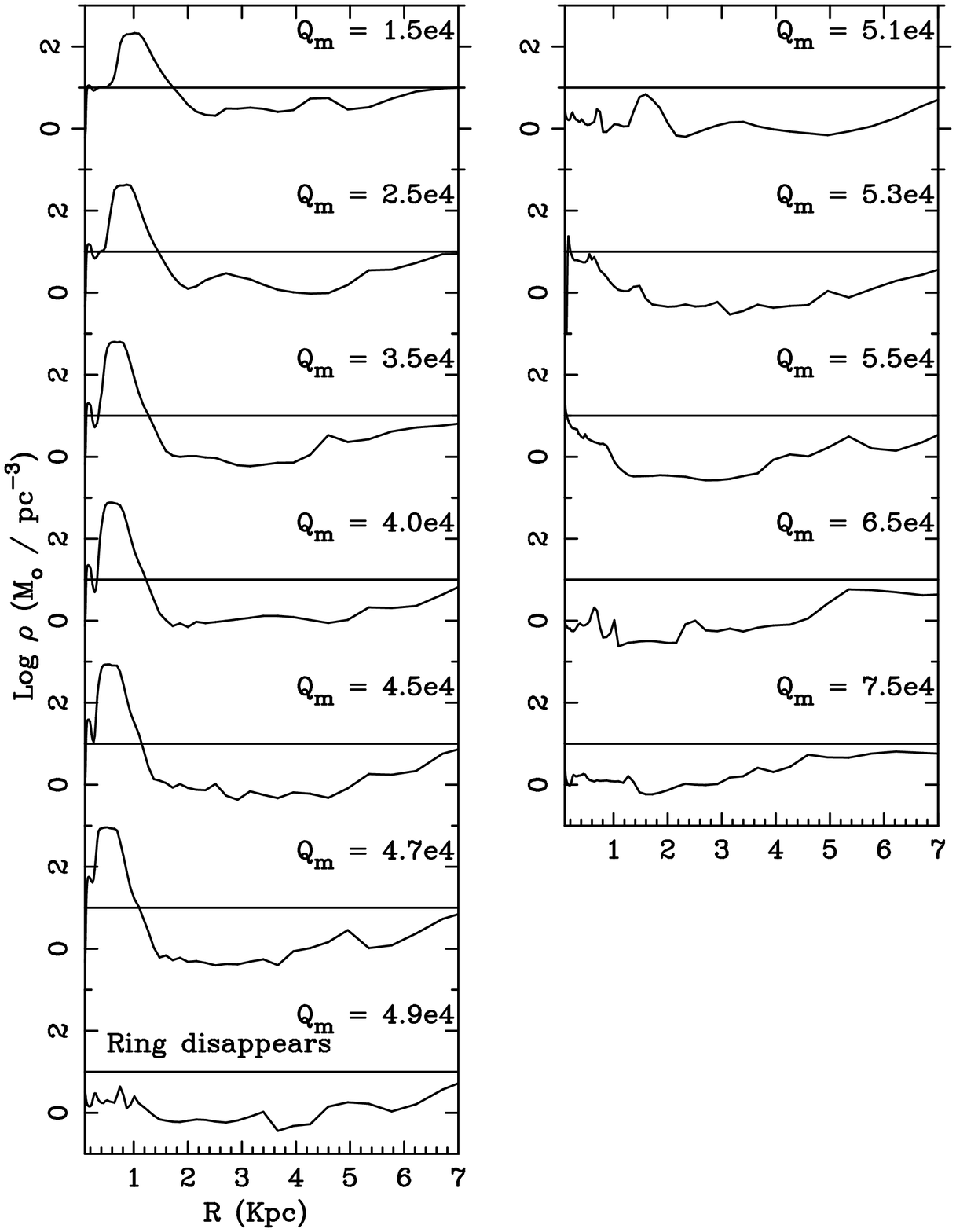,width=3.0in,angle=0,silent=1}
 \vskip 0.2in
 \psfig{figure=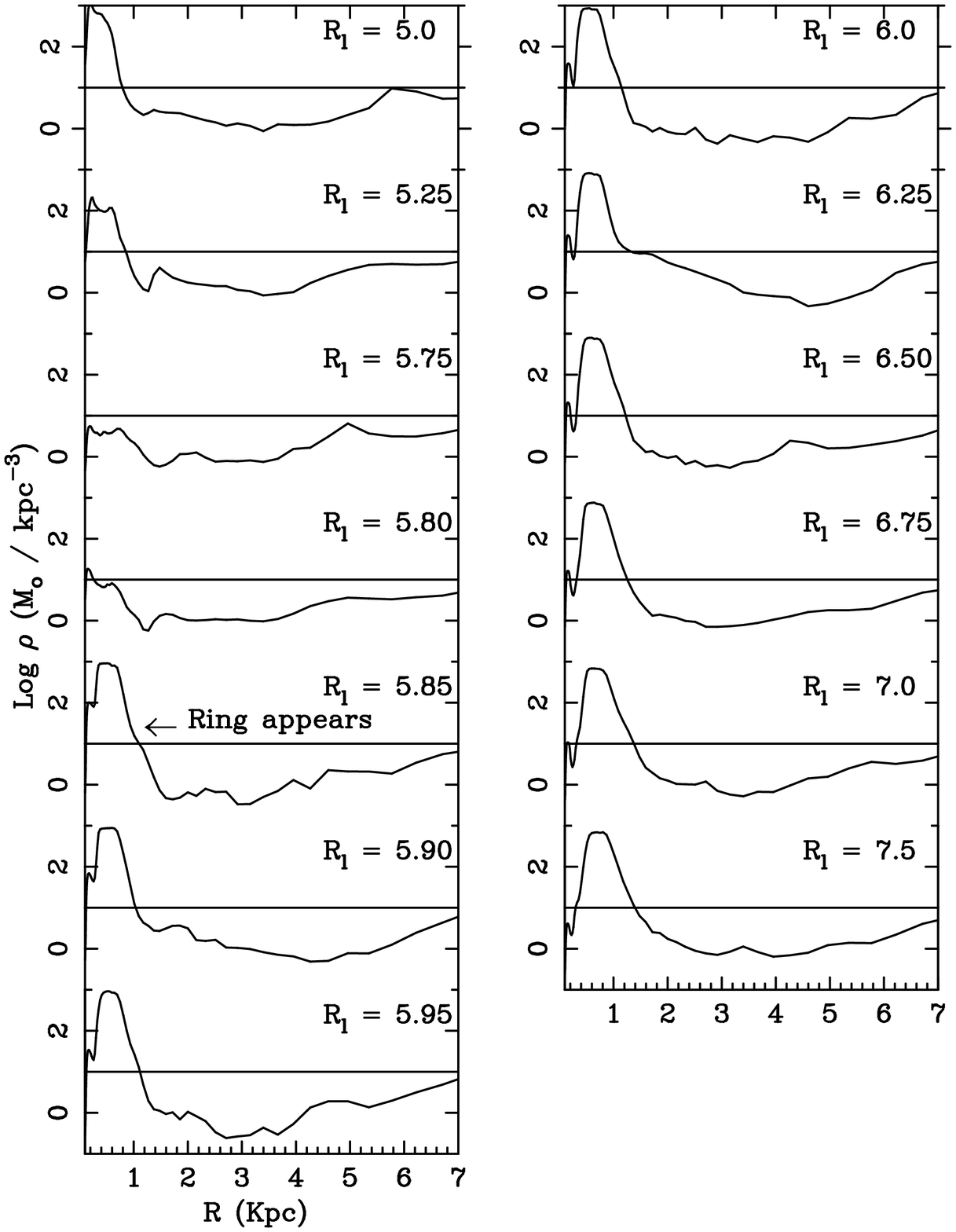,width=3.0in,angle=0,silent=1}

\caption[Models]
{Azimuthally averaged log density versus radius for different
  hydrodynamic models.  The horizontal line is the initial density.
  The standard parameters are a/b=4.0, $\rho_c$ = 2.4e4 x 10$^6$ \Msun
  kpc$^{-3}$, Q$_m$=4.5e4 x 10$^6$ \Msun kpc$^{2}$, r$_l$=6.0 kpc.
  When one of the three free parameters is varied (as indicated on the
  individual plots), the others are set to these standard
  values.\label{modfig}}
\end{figure}

The PST95 hydrodynamic models are computed with a third-order
piecewise, parabolic method (PPM) with a cylindrical coordinate system
yielding a resolution of 8 pc at a distance of 0.1 kpc.   
The gravitational potential used in this model consists of a
Kuzmin-Toomre disk, a bulge and a bar described by a Ferrer's
ellipsoid.  
The model is completely determined by four independent input parameters:
the bar axial ratio (a/b), the central density ($\rho_c$, 10$^6$ \Msun
kpc$^{-3}$ ), the Lagrangian radius (r$_l$, kpc, equivalently the
pattern speed), and the quadrupole moment (q$_m$, 10$^6$ \Msun
kpc$^{2}$, equivalently the bar mass) (see PST95 for details).  
We chose to set a/b = 4.0 and varied the other three parameters as
follows: 1.0e4 $<$ $\rho_c$  $<$ 7.4e4, 5.0 $<$ r$_l$ $<$ 7.5, 1.5e4
$<$ q$_m$ $<$ 7.5e4.  
Since any of these three free parameters can affect the transition
between the two outcomes of the model, we varied one parameter at a
time while setting the other two parameters to the ``standard'' model
values used by PST95.  
As in PST95, we ran each simulation until the bar had settled into a
quasi-static state i.e. until t = 2.0 Gyr (see PST95 for more details
on time evolution of the bar).  
The final azimuthally averaged log density versus radius are plotted
in Figure \ref{modfig}.  

We find that the formation of the ring is indeed abrupt in all
instances where an ILR exists.  
In other words, if offset dust lanes form, a nuclear ring inevitably forms.  
With this result, we conclude that the intermediate case, if present,
exists in even finer steps of the model parameters; otherwise we
confirm PST95's assertions about the robustness of the nuclear ring.  
Although this experiment failed in finding the intermediate scenario
which may have explained the observations of NGC~5383, we discovered 
other interesting results which add to our current understanding of
gas distribution in the hydrodynamic models.

\begin{figure}[!htb]
\centerline{\psfig{figure=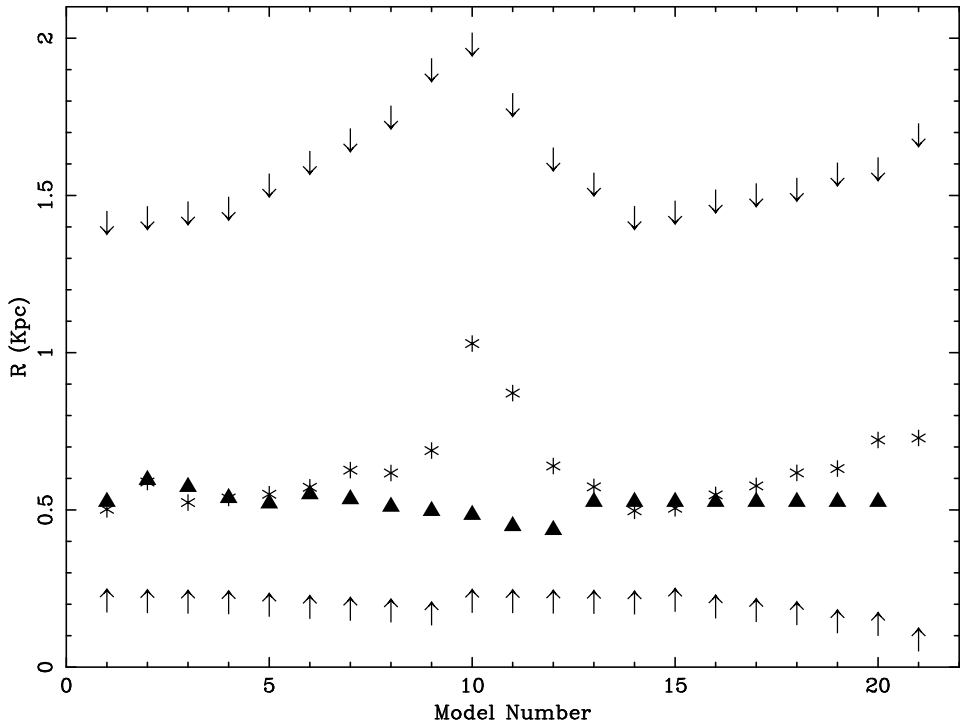,angle=-90,width=5in}}
\caption[Location of ring]
{A plot of the Lindblad numbers: OILR (down arrow), IILR (up arrow),
  maximum (triangle) of the $\Omega$-$\kappa$/2 curve and the ring
  peak (star) for different models (Table \ref{models}) is shown.
  The ring peak always lies between the OILR and the IILR.  Its
  location is not coincident with the maximum of the
  $\Omega$-$\kappa$/2 curve but seems to be determined by the location
  of  the OILR.\label{ilrs}}
\end{figure}

First, when the ring exists, we find that the ring seems to be
located at approximately the same distance and has approximately the
same thickness growing only slightly thinner with higher central
density concentration.  
Since the commonly accepted explanation for the location of the
nuclear ring is that it must exist between the outer ILR (OILR) and
the inner ILR (IILR), one can conclude that the ring location and
thickness should be restricted to a certain range of radii.  
We find that in all 21 model runs (Table \ref{models}) which produce a
ring, the ring peak is always located between the OILR and the IILR as
shown in Figure \ref{ilrs}.   

Interestingly, we also find that the location of the ring peak does
not coincide with the location of the $\Omega$-$\kappa$/2 maximum
(Figure \ref{ilrs}) as suggested by PST95 except for the standard model.
In most cases the ring peak is close to the maximum of the
$\Omega$-$\kappa$/2 curve, but its location seems to be determined by
the location of the OILR for a given model.  
In other words the location of the ring peak and the OILR varies in
exactly the same way for the different models but by different amounts from
model to model.  
While this relationship is interesting, we note that it may not be
physical since the OILR is located at a large distance ($\sim$ 1.5-2
kpc) from the center; at these distance, the gas is no longer on
circular orbits and hence the classical definition of an OILR may not be
relevant.  
It would be, perhaps, more interesting to compare the location of the
ring peak to the x1 and x2 orbits in the various models.

\end{document}